# Massless Wave States of Two-Fermion Systems

A.I. Agafonov

Russian Research Center"Kurchatov Institute"

Kurchatov sq. 1, Moscow 123182, Russia

E-mail: aai@isssph.kiae.ru

**Abstract**

It is known that in the ladder approximation the relativistic two-fermion bound-state equation of Bethe and Salpeter has solutions corresponding to the binding energy equal to the total mass of the particles. The study of these massless states has been carried out only for the bound system at rest. Of course, such composite boson can not be in the state of rest. But it is more importantly that this approach for the massless boson can not be interpreted as the limiting case of a nonzero mass system because the phase velocity of the boson wave must equal to the speed of light.

Using the Bethe-Salpeter equation in the ladder approximation, we have obtained the wave equation for the massless bound states of two fermions with equal masses and the electromagnetic interaction between them. Neglecting retardation of the interaction, solutions corresponded to the stationary wave states of the composite boson, have been found. The boson wave function can be represented as an infinite, straight thread, the transverse radius of which is more than the Compton wavelength of the fermion. Two energy branches of the bosons with quantized energies have been determined. The appearance of these branches is due to the strong renormalization of the fine structure constant for the massless states.

**1. Intoduction**

For the fist time massless bound states for the two-fermion system with equal masses have been obtained in [1] by solving the Bethe-Salpeter equation (BSE) in the ladder approximation. The binding energy of the composite boson in this strongly-coupled bound state equals to $2m$, where $m$ is the fermion mass. If the interaction between the fermions takes place by means of massless quanta, then these solutions were shown to arise for any values of coupling constants. The ladder approximation was justified in [2] for small coupling constants, in particular, for the electromagnetic interaction.

It should be noted that in [1] and following works [3-10] concerning this problem, massless states have been studied only for the bound system at rest. However, such zero mass composite particles can not be in a state of rest. In a sense, its state must be like as the photon one. Here we refer to the momentum dependence of the boson energy but certainly not the kind of the wave function.

It is known that the BSE can have normal and abnormal solutions [11,12]. For the normal solutions the binding energy vanishes when the coupling constant tends to zero. However, it remains unclear whether the abnormal solutions for which the binding energy vanishes at finite values of the coupling constant, should be considered as non-physical [13]. As noted above, the Goldstein solutions are not associated with any threshold value of the coupling constant (see, also, [14] and references therein). This is encouraging that these strongly bound states may be physical.

In the present paper we search the BSE solution corresponding to the massless wave states of the two-fermion systems. Results presented below are obtained with the neglecting of the interaction retardation. Therefore they should be regarded only as a prediction of the existence of the massless wave states of the composite bosons.

Experiments can only either confirm or reject the existence of these massless bosons. A test is proposed in which these particles, if they exist, could be detected.

**2. Bound-state equation**

We consider a system of two non-identical spin-halt particles with equal masses, charges $\pm e$ and the electromagnetic interaction between them. Henceforth operators are introduced the lower index $\pm$ showing their action on functions of particles with negative (-) and positive (+) charge, respectively.

The free particle propagator is determined by the differential equation:

$$(\gamma \hat{p}_2 - m) K(2,1) = \delta^4(2,1), \qquad (1)$$

where $\gamma_\mu$ - the Dirac matrices in the standard representation, the low index of the operator $\hat{p}_\mu = i\partial_\mu$ means its action on the 4-radius-vector $x_{2\mu}$ of the propagation $K(2,1)$ (system of units $\hbar = c = 1$ is used). The propagator satisfying (1), can be written as (see [15]):

$$K(2,1) = -i\sum_{\mathbf{p}}\left\{\psi_p(\mathbf{x}_2)\overline{\psi}_p(\mathbf{x}_1)e^{-i\varepsilon_\mathbf{p}(t_2-t_1)} + \psi_{-p}(\mathbf{x}_2)\overline{\psi}_{-p}(\mathbf{x}_1)e^{+i\varepsilon_\mathbf{p}(t_2-t_1)}\right\}\theta(t_2-t_1). \quad (2)$$

Taking into account the interaction function between the particles [2,14], it is convenient to introduce the matrixes $\gamma_\pm^0$ in the definition of $K_\pm(2,1)$ (2). Using the expressions of the plane wave functions $\psi_{\pm p}$, the temporal Fourier components of the propagators are given by:

$$K_-(\mathbf{r}_2 - \mathbf{r}_1;\omega) = \sum_{\mathbf{p}}\left\{\frac{\Lambda_-^+(\mathbf{p})}{\omega - \varepsilon_\mathbf{p} + i\delta} + \frac{\Lambda_-^-(\mathbf{p})}{\omega + \varepsilon_\mathbf{p} + i\delta}\right\}\exp[i\mathbf{p}(\mathbf{r}_2-\mathbf{r}_1)] \quad (3)$$

and

$$K_+(\mathbf{r}_2 - \mathbf{r}_1;\omega) = \sum_{\mathbf{q}}\left\{\frac{\Lambda_+^+(\mathbf{q})}{\omega - \varepsilon_\mathbf{q} + i\delta} + \frac{\Lambda_+^-(\mathbf{q})}{\omega + \varepsilon_\mathbf{q} + i\delta}\right\}\exp[i\mathbf{q}(\mathbf{r}_2-\mathbf{r}_1)], \quad (4)$$

where $\delta \to 0^+$, and

$$\Lambda_-^\pm(\mathbf{p}) = \frac{\varepsilon_\mathbf{p} I_- \pm \boldsymbol{\alpha}_-\mathbf{p} \pm m\gamma_-^0}{2\varepsilon_\mathbf{p}}, \quad \Lambda_+^\pm(\mathbf{q}) = \frac{\varepsilon_\mathbf{q} I_+ \pm \boldsymbol{\alpha}_+\mathbf{q} \pm m\gamma_+^0}{2\varepsilon_\mathbf{q}}.$$

Here $I_\pm$ are the unit matrixes, $\boldsymbol{\alpha}_\pm$ are $\boldsymbol{\alpha}$-matrixes in the standard representation.

In the ladder approximation the interaction function between the fermions is [2]:

$$G^{(1)}(1,2;3,4) = -e^2 \frac{1-\boldsymbol{\alpha}_-\boldsymbol{\alpha}_+}{|\mathbf{r}_1-\mathbf{r}_2|}\int_{-\infty}^{+\infty}\frac{d\omega}{2\pi}\exp[-i\omega(t_1-t_2)+i|\omega||\mathbf{r}_1-\mathbf{r}_2|]\delta^4(1,3)\delta^4(2,4). \quad (5)$$

Using (5), the BSE for the wave function of bound states is

$$\psi(\mathbf{r}_3,t_3;\mathbf{r}_4,t_4) = -ie^2 \int d\mathbf{r}_5 \int dt_5 \int d\mathbf{r}_6 \int dt_6 \int d\mathbf{r}_7 \int dt_7 \int d\mathbf{r}_8 \int dt_8 \frac{d\omega_1}{2\pi}\int\frac{d\omega_2}{2\pi}\int\frac{d\omega_3}{2\pi}$$

$$K_-(\mathbf{r}_3-\mathbf{r}_5;\omega_1)K_+(\mathbf{r}_4-\mathbf{r}_6;\omega_2)\frac{1-\boldsymbol{\alpha}_-\boldsymbol{\alpha}_+}{|\mathbf{r}_5-\mathbf{r}_6|}$$

$$\exp[-i\omega_1(t_3-t_5) - i\omega_2(t_4-t_6) - i\omega_3(t_5-t_6) + i|\omega_3||\mathbf{r}_5-\mathbf{r}_6|]$$

$$\theta(t_3-t_5)\theta(t_4-t_6)\delta(t_5-t_7)\delta(\mathbf{r}_5-\mathbf{r}_7)\delta(t_6-t_8)\delta(\mathbf{r}_6-\mathbf{r}_8)\psi(\mathbf{r}_7,t_7;\mathbf{r}_8,t_8). \quad (6)$$

Integrating over $\delta$-functions in Eq. (6), we have:

$$\psi(\mathbf{r}_3,t_3;\mathbf{r}_4,t_4) = -ie^2 \int d\mathbf{r}_7 \int_{-\infty}^{t_3} dt_7 \int d\mathbf{r}_8 \int_{-\infty}^{t_4} dt_8 \int_{-\infty}^{+\infty} \frac{d\omega_1}{2\pi} \int_{-\infty}^{+\infty} \frac{d\omega_2}{2\pi} \int_{-\infty}^{+\infty} \frac{d\omega_3}{2\pi}$$

$$K_-(\mathbf{r}_3 - \mathbf{r}_7;\omega_1) K_+(\mathbf{r}_4 - \mathbf{r}_8;\omega_2) \frac{1 - \boldsymbol{\alpha}_- \boldsymbol{\alpha}_+}{|\mathbf{r}_7 - \mathbf{r}_8|}$$

$$\exp(-i\omega_1(t_3 - t_7) - i\omega_2(t_4 - t_8) - i\omega_3(t_7 - t_8) + i|\omega_3||\mathbf{r}_7 - \mathbf{r}_8|) \psi(\mathbf{r}_7,t_7;\mathbf{r}_8,t_8). \qquad (7)$$

It seems that any relativistic theories inhere in states with negative energies. Therefore, along with the positive energy states of the composite boson, the case of negative energies should also be explored. For waves with positive energies the particle moves forward in time, where as solutions of Eq. (7) with negative energies should be understood as the moving backward with positive energies also. For the latter case in Eq. (6) it should make the replacement of two $\theta$ - functions $\theta(t_3 - t_5)\theta(t_4 - t_6)$ on $\theta(t_5 - t_3)\theta(t_6 - t_4)$. As a result, Eq. (7) is transformed as:

$$\psi(\mathbf{r}_3,t_3;\mathbf{r}_4,t_4) = -ie^2 \int d\mathbf{r}_7 \int_{t_3}^{+\infty} dt_7 \int d\mathbf{r}_8 \int_{t_4}^{+\infty} dt_8 \int_{-\infty}^{+\infty} \frac{d\omega_1}{2\pi} \int_{-\infty}^{+\infty} \frac{d\omega_2}{2\pi} \int_{-\infty}^{+\infty} \frac{d\omega_3}{2\pi}$$

$$K_-(\mathbf{r}_3 - \mathbf{r}_7;\omega_1) K_+(\mathbf{r}_4 - \mathbf{r}_8;\omega_2) \frac{1 - \boldsymbol{\alpha}_- \boldsymbol{\alpha}_+}{|\mathbf{r}_7 - \mathbf{r}_8|}$$

$$\exp(-i\omega_1(t_3 - t_7) - i\omega_2(t_4 - t_8) - i\omega_3(t_7 - t_8) + i|\omega_3||\mathbf{r}_7 - \mathbf{r}_8|) \psi(\mathbf{r}_7,t_7;\mathbf{r}_8,t_8). \qquad (8)$$

There is no need to separately investigate Eq. (8). One can verify that it is enough to change the sign of the wave energy in the problem on eigenfunctions and eigenvalues obtained from Eq. (7).

### 3. Bound states without retardation of the interaction

In this case it should be put $|\omega_3||\mathbf{r}_7 - \mathbf{r}_8| = 0$ in Eq. (7). Then the integral over $\omega_3$ in (7) yields $\delta(t_7 - t_8)$ and, consequently, the wave function of the bound pair is dependent on single time for the particles $t_3 = t_4$. Because the solution of Eq. (7) is sought for the stationary wave states of the composite boson, its wave function can be represented as:

$$\psi(\mathbf{r}_3,t_3;\mathbf{r}_4,t_3) = \varphi(\mathbf{r}_3,\mathbf{r}_4)\exp(-i\omega t_3), \qquad (9)$$

where $\omega$ - the boson energy. Substituting (9) in Eq. (7) and integrating over $t_7$ and $t_8$, we obtain:

$$\varphi(\mathbf{r}_3,\mathbf{r}_4) = -ie^2 \int d\mathbf{r}_7 \int d\mathbf{r}_8 \int_{-\infty}^{+\infty} \frac{d\omega_1}{2\pi} \int_{-\infty}^{+\infty} d\omega_2 \delta(\omega - \omega_1 - \omega_2)$$

$$K_-(\mathbf{r}_3 - \mathbf{r}_7;\omega_1) K_+(\mathbf{r}_4 - \mathbf{r}_8;\omega_2) \frac{1 - \boldsymbol{\alpha}_-\boldsymbol{\alpha}_+}{|\mathbf{r}_7 - \mathbf{r}_8|} \varphi(\mathbf{r}_7,\mathbf{r}_8). \quad (10)$$

Taking into account (3) and (4), in Eq. (10) the contour of integration over $\omega_2$ is closed in the upper half plan $\text{Re}\,\omega_1 > 0$ where there are two simple poles $\omega_1 = \omega \pm \varepsilon_\mathbf{q} + i\delta$. As a result, we have:

$$\varphi(\mathbf{r}_3,\mathbf{r}_4) = -e^2 \int d\mathbf{r}_7 \int d\mathbf{r}_8 K_{-+}(\mathbf{r}_3 - \mathbf{r}_7, \mathbf{r}_4 - \mathbf{r}_8;\omega) \frac{1 - \boldsymbol{\alpha}_-\boldsymbol{\alpha}_+}{|\mathbf{r}_7 - \mathbf{r}_8|} \varphi(\mathbf{r}_7,\mathbf{r}_8). \quad (11)$$

Here the propagation function $K_{-+}(\mathbf{r}_3 - \mathbf{r}_7, \mathbf{r}_4 - \mathbf{r}_8;\omega)$ is:

$$K_{-+}(\mathbf{r}_3 - \mathbf{r}_7, \mathbf{r}_4 - \mathbf{r}_8;\omega) = \int \frac{d\mathbf{p}}{(2\pi)^3} \int \frac{d\mathbf{q}}{(2\pi)^3} K_{-+}(\mathbf{p},\mathbf{q};\omega) \exp[i\mathbf{p}(\mathbf{r}_3 - \mathbf{r}_7) + i\mathbf{q}(\mathbf{r}_4 - \mathbf{r}_8)], \quad (12)$$

where

$$K_{-+}(\mathbf{p},\mathbf{q};\omega) = \frac{\Lambda_-^+(p)\Lambda_+^+(q)}{\omega - \varepsilon_\mathbf{p} - \varepsilon_\mathbf{q}} + \frac{\Lambda_-^-(p)\Lambda_+^+(q)}{\omega + \varepsilon_\mathbf{p} - \varepsilon_\mathbf{q}} + \frac{\Lambda_-^+(p)\Lambda_+^-(q)}{\omega - \varepsilon_\mathbf{p} + \varepsilon_\mathbf{q}} + \frac{\Lambda_-^-(p)\Lambda_+^-(q)}{\omega + \varepsilon_\mathbf{p} + \varepsilon_\mathbf{q}}. \quad (13)$$

If the terms witn the negative energies are neglected, then (13) becomes the Logunov-Tavkhelidze propagator [16]. In the case of neglecting the second and third terms on the right side of (13) (the cross terms with positive and negative energies), one obtains the propagator commonly used in the study of bound states the two-fermion systems.

The right side of (13) is conveniently reduced to a common denominator. As a result one obtains:

$$K_{-+}(\mathbf{p},\mathbf{q};\omega) = \frac{1}{\omega^4 - 2\omega^2(\varepsilon_p^2 + \varepsilon_q^2) + (\varepsilon_p^2 - \varepsilon_q^2)^2} \Big\{\omega(\omega^2 - \varepsilon_p^2 - \varepsilon_q^2) I_- I_+ +$$

$$+ 2m^2 \omega \gamma_-^0 \gamma_+^0 + (\omega^2 - \varepsilon_p^2 + \varepsilon_q^2)(\boldsymbol{\alpha}_-\mathbf{p}) I_+ + (\omega^2 + \varepsilon_p^2 - \varepsilon_q^2)(\boldsymbol{\alpha}_+\mathbf{q}) I_- + 2\omega(\boldsymbol{\alpha}_-\mathbf{p})(\boldsymbol{\alpha}_+\mathbf{q}) +$$

$$m(\omega^2 - \varepsilon_p^2 + \varepsilon_q^2)\gamma_-^0 I_+ + m(\omega^2 + \varepsilon_p^2 - \varepsilon_q^2)\gamma_+^0 I_- + 2m\omega(\gamma_-^0(\boldsymbol{\alpha}_+\mathbf{q}) + \gamma_+^0(\boldsymbol{\alpha}_-\mathbf{p}))\Big\}. \quad (14)$$

Note that the sequence of the operators with subscripts $\pm$ is arbitrary since they commute. For example, the term with $(\boldsymbol{\alpha}_-\mathbf{p})(\boldsymbol{\alpha}_+\mathbf{q})$ in the right side of Eq. (14) can equally be written in the form $(\boldsymbol{\alpha}_+\mathbf{q})(\boldsymbol{\alpha}_-\mathbf{p})$ because the matrixes $\alpha_{+\mu}$ and $\alpha_{-\mu}$ operate on different spin variables of the composite particle wave function. These wave functions are presented below.

Using (14), is easy to see that the integral equation (11) is corresponded to the following differential equation on the eigenfunctions $\varphi$ and eigenvalues $\omega$ of the bound pair:

$$\left[\omega^4 - 2\omega^2(\varepsilon_p^2+\varepsilon_q^2)+(\varepsilon_p^2-\varepsilon_q^2)^2\right]\varphi(\mathbf{r}_1,\mathbf{r}_2) = -e^2\Big\{\omega(\omega^2-\varepsilon_p^2-\varepsilon_q^2)I_-I_+ + 2\omega(\boldsymbol{\alpha}_-\mathbf{p})(\boldsymbol{\alpha}_+\mathbf{q})+$$

$$(\omega^2-\varepsilon_p^2+\varepsilon_q^2)(\boldsymbol{\alpha}_-\mathbf{p})I_+ + (\omega^2+\varepsilon_p^2-\varepsilon_q^2)(\boldsymbol{\alpha}_+\mathbf{q})I_- + 2m\omega(\gamma_-^0(\boldsymbol{\alpha}_+\mathbf{q})+\gamma_+^0(\boldsymbol{\alpha}_-\mathbf{p}))+$$

$$m(\omega^2-\varepsilon_p^2+\varepsilon_q^2)\gamma_-^0 I_+ + m(\omega^2+\varepsilon_p^2-\varepsilon_q^2)\gamma_+^0 I_- + 2m^2\omega\gamma_-^0\gamma_+^0\Big\}\frac{1-\boldsymbol{\alpha}_-\boldsymbol{\alpha}_+}{|\mathbf{r}_1-\mathbf{r}_2|}\varphi. \quad (15)$$

Here $\hat{\mathbf{p}} = -i\nabla_{\mathbf{r}_1}$ and $\hat{\mathbf{q}} = -i\nabla_{\mathbf{r}_2}$ are canonical momentums.

**4. Analysis of Eq. (15) for the system at rest**

In the case $\mathbf{p}+\mathbf{q}=0$. Introducing the relative radius-vector $\mathbf{r}=\mathbf{r}_1-\mathbf{r}_2$, Eq. (15) is reduced to the form:

$$\omega^2\left[\omega^2-4\varepsilon_p^2\right]\varphi = -e^2\omega\Big\{\!\!\left(\omega^2-2\varepsilon_p^2\right)\!I_-I_+ + \omega[(\boldsymbol{\alpha}_-\mathbf{p})I_+ - (\boldsymbol{\alpha}_+\mathbf{p})I_-] + m\omega\left(\gamma_-^0 I_+ + \gamma_+^0 I_-\right)$$

$$- 2(\boldsymbol{\alpha}_-\mathbf{p})(\boldsymbol{\alpha}_+\mathbf{p}) + 2m\left[\gamma_+^0(\boldsymbol{\alpha}_-\mathbf{p}) - \gamma_-^0(\boldsymbol{\alpha}_+\mathbf{p})\right] + 2m^2\gamma_-^0\gamma_+^0\Big\}(1-\boldsymbol{\alpha}_-\boldsymbol{\alpha}_+)\frac{1}{r}\varphi(\mathbf{r}). \quad (16)$$

Here the operator $\mathbf{p} = -i\nabla_{\mathbf{r}}$.

If the energy $\omega = 0$ then the left and right sides of Eq. (16) vanish. However, this energy eigenvalue would correspond to the zero-mass composite boson, and the binding energy in this state would be equal the total mass of the particles. Then, this boson can not remain at rest and its state would be described by a wave with the phase velocity equal to the speed of light. Therefore, Eq. (16) shows only the possibility of such state. Its wave function should be sought from Eq. (11) in accordance with its energy $\omega^2 = g^2$, where the momentum of the bound pair $\mathbf{g} = \mathbf{p} + \mathbf{q}$ is nonzero. This will be done in the next section.

One can verify that Eq. (16) has solutions for loosely bound states. For these states the energy of the system $\omega = 2m + E$, where the bound energy $E \propto -\alpha^2 m$, the characteristic value of the momentum $p \propto \alpha m$ and the matrix elements of $\boldsymbol{\alpha}_\pm \propto \alpha$ ($\alpha$ - the fine structure constant).

## 5. Equation of the massless boson

We are looking for the solutions of Eq. (11) for which the bound pair is represented a wave with the phase velocity equal to the speed of light. In general case the energy of the wave $\omega = \pm g$, where $\mathbf{g} = \mathbf{p} + \mathbf{q}$ is its momentum.

Let the momentum $\mathbf{g}$ is parallel to the axis $z$. In the coordinate part of the wave function it is convenient to separate the motion of the bound particles as a whole from their relative motion. For the considered wave the z-components of the radius - vectors of the two particles must be equal, $z_3 = z_4$. Then the relative radius - vector between the particles will be two-dimensional, $\boldsymbol{\rho}(x,y) = \boldsymbol{\rho}_3 - \boldsymbol{\rho}_4$. Therefore, the wave function of the state is sought in the form:

$$\varphi(\mathbf{r}_3, \mathbf{r}_4) = \varphi(\boldsymbol{\rho}_3 - \boldsymbol{\rho}_4; g) \exp\left[i\frac{g}{2}(z_3 + z_4)\right] \qquad (17)$$

at the condition $z_3 = z_4$. The function $\varphi(\mathbf{r}_7, \mathbf{r}_8)$ in Eq. (11) has the same representation (17) with the replacement of the subscripts 3,4 to 7,8.

According to (17), the transverse motion of the coupled fermions has a two-dimensional character, and is determined by the wave function $\varphi(\boldsymbol{\rho}; g)$. The wave function (17) can be represented as an infinite, straight thread with the transverse radius determined by $\varphi(\boldsymbol{\rho}; g)$.

In Eq. (11) it is replaced the integration variables: $d\mathbf{r}_7 d\mathbf{r}_8 = d\boldsymbol{\rho}_1 d\mathbf{R}_1 dz_7 dz_8$ where $\boldsymbol{\rho}_1 = \boldsymbol{\rho}_7 - \boldsymbol{\rho}_8$ and $2\mathbf{R}_1 = \boldsymbol{\rho}_7 + \boldsymbol{\rho}_8$. Respectively, instead of $\boldsymbol{\rho}_3, \boldsymbol{\rho}_4$ it is introduced the variables $\boldsymbol{\rho} = \boldsymbol{\rho}_3 - \boldsymbol{\rho}_4$ and $2\mathbf{R} = \boldsymbol{\rho}_3 + \boldsymbol{\rho}_4$. Substituting the function (17) in Eq. (11), the integration over the variables $z_7, z_8$ leads to $(2\pi)^2 \delta(p_z - \frac{1}{2}g)\delta(q_z - \frac{1}{2}g)$, and the integration over $\mathbf{R}_1$ results in $(2\pi)^2 \delta(\mathbf{p}_\perp + \mathbf{q}_\perp)$. Then it is easily carried out the integration over $p_z, q_z$ and $\mathbf{q}_\perp$. As a result, we obtain the integral equation:

$$\varphi(\boldsymbol{\rho}) = -\alpha \int d\boldsymbol{\rho}_1 \int \frac{d\mathbf{p}_\perp}{(2\pi)^2} e^{i\mathbf{p}_\perp(\boldsymbol{\rho}-\boldsymbol{\rho}_1)} K_{-+}(\mathbf{p}_\perp + \frac{\mathbf{g}}{2}, -\mathbf{p}_\perp + \frac{\mathbf{g}}{2}; \omega) \frac{1 - \boldsymbol{\alpha}_- \boldsymbol{\alpha}_+}{\rho_1} \varphi(\boldsymbol{\rho}_1), \quad (18)$$

where $\alpha$ - the fine structure constant.

The rearrangement of the spatial coordinates of the fermions means the replacement $\boldsymbol{\rho} \to -\boldsymbol{\rho}$ and $\boldsymbol{\rho}_1 \to -\boldsymbol{\rho}_1$. Making the replacement of the integration variable $\mathbf{p}_\perp \to -\mathbf{p}_\perp$, Eq. (18) is transformed to the form:

$$\varphi(-\boldsymbol{\rho}) = -\alpha \int d\boldsymbol{\rho}_1 \int \frac{d\mathbf{p}_\perp}{(2\pi)^2} e^{i\mathbf{p}_\perp(\boldsymbol{\rho}-\boldsymbol{\rho}_1)} K_{-+}(-\mathbf{p}_\perp + \frac{\mathbf{g}}{2}, \mathbf{p}_\perp + \frac{\mathbf{g}}{2}; \omega) \frac{1 - \boldsymbol{\alpha}_- \boldsymbol{\alpha}_+}{\rho_1} \varphi(-\boldsymbol{\rho}_1).$$

The wave function of the transverse motion is transformed as $\varphi(-\boldsymbol{\rho}) = \pm \varphi(\boldsymbol{\rho})$ according to the spatial parity of the state. In both cases of the wave function symmetry we obtain Eq. (18) with the replacement:

$$K_{-+} \to \tilde{K}_{-+} = \frac{1}{2}\left[ K_{-+}(\mathbf{p}_\perp + \frac{\mathbf{g}}{2}, -\mathbf{p}_\perp + \frac{\mathbf{g}}{2}; \omega) + K_{-+}(-\mathbf{p}_\perp + \frac{\mathbf{g}}{2}, \mathbf{p}_\perp + \frac{\mathbf{g}}{2}; \omega) \right]$$

As a results, Eq. (18) is rewritten as:

$$\varphi(\boldsymbol{\rho}) = -\frac{\alpha}{\omega} \int d\boldsymbol{\rho}_1 \int \frac{d\mathbf{p}_\perp}{(2\pi)^2} \frac{e^{i\mathbf{p}_\perp(\boldsymbol{\rho}-\boldsymbol{\rho}_1)}}{\omega^2 - 4\varepsilon_p^2} \Big\{ (\omega^2 - 2\varepsilon_p^2) + m\omega(\gamma_-^0 + \gamma_+^0) + \frac{1}{2}g\omega(\alpha_-^z + \alpha_+^z) +$$

$$\frac{1}{2}g^2 \alpha_-^z \alpha_+^z - 2\Big[p_x^2 \alpha_-^x \alpha_+^x + p_y^2 \alpha_-^y \alpha_+^y + \hat{p}_x \hat{p}_y (\alpha_-^x \alpha_+^y + \alpha_-^y \alpha_+^x)\Big] +$$

$$mg(\alpha_-^z \gamma_+^0 + \alpha_+^z \gamma_-^0) + 2m^2 \gamma_-^0 \gamma_+^0 \Big\} (1 - \boldsymbol{\alpha}_- \boldsymbol{\alpha}_+) \frac{1}{\rho_1} \varphi(\boldsymbol{\rho}_1; g). \quad (19)$$

Note that, in general, in Eq. (19) there are two possibilities $\omega = \pm g$, corresponding to the positive and negative energy of the waves. In the latter case, as noted above, it should be understood as the moving of the boson backward in time with positive energies also.

### 6. The spin states

The composite boson can be in four different spin states corresponding to the eigenvalues of the square of the total spin:

$$\mathbf{S}^2 = \frac{1}{2}\Big[3 + \Sigma_-^x \Sigma_+^x + \Sigma_-^y \Sigma_+^y + \Sigma_-^z \Sigma_+^z\Big]$$

and its projections on the momentum $\mathbf{g}$ that is parallel to the z-axis:

$$\hat{S}_z = \frac{1}{2}\Big[\Sigma_-^z + \Sigma_+^z\Big],$$

where the matrixes $\boldsymbol{\Sigma}_{\pm} = \begin{pmatrix} \boldsymbol{\sigma}_{\pm} & 0 \\ 0 & \boldsymbol{\sigma}_{\pm} \end{pmatrix}$.

The first subscript of the wave functions presented below, corresponds to the quantum number of the total spin ($S=1$ for the triplet state or $S=0$ for the singlet state), and the second index - the eigenvalue of $S_z : \varphi_{SS_z}$. In the following wave functions the first bispinor is referred to the negatively charged particle.

The singlet wave function $\varphi_{00}$ of the transverse motion of the coupled fermions is:

$$\varphi_{00} = \frac{1}{\sqrt{8}} \left\{ \theta_1(\boldsymbol{\rho},\mathbf{g}) \left[ \begin{pmatrix} 1 \\ 0 \\ 0 \\ 0 \end{pmatrix}\begin{pmatrix} 0 \\ 1 \\ 0 \\ 0 \end{pmatrix} - \begin{pmatrix} 0 \\ 1 \\ 0 \\ 0 \end{pmatrix}\begin{pmatrix} 1 \\ 0 \\ 0 \\ 0 \end{pmatrix} \right] + \theta_2(\boldsymbol{\rho},\mathbf{g}) \left[ \begin{pmatrix} 0 \\ 0 \\ 1 \\ 0 \end{pmatrix}\begin{pmatrix} 0 \\ 0 \\ 0 \\ 1 \end{pmatrix} - \begin{pmatrix} 0 \\ 0 \\ 0 \\ 1 \end{pmatrix}\begin{pmatrix} 0 \\ 0 \\ 1 \\ 0 \end{pmatrix} \right] + \right.$$

$$\left. \theta_3(\boldsymbol{\rho},\mathbf{g}) \left[ \begin{pmatrix} 0 \\ 0 \\ 1 \\ 0 \end{pmatrix}\begin{pmatrix} 0 \\ 1 \\ 0 \\ 0 \end{pmatrix} - \begin{pmatrix} 0 \\ 1 \\ 0 \\ 0 \end{pmatrix}\begin{pmatrix} 0 \\ 0 \\ 1 \\ 0 \end{pmatrix} \right] + \theta_4(\boldsymbol{\rho},\mathbf{g}) \left[ \begin{pmatrix} 1 \\ 0 \\ 0 \\ 0 \end{pmatrix}\begin{pmatrix} 0 \\ 0 \\ 0 \\ 1 \end{pmatrix} - \begin{pmatrix} 0 \\ 0 \\ 0 \\ 1 \end{pmatrix}\begin{pmatrix} 1 \\ 0 \\ 0 \\ 0 \end{pmatrix} \right] \right\}. \quad (20)$$

The functions $\theta_i$ must be normalized: $\int |\theta_i|^2 \rho d\rho d\phi = 1$.

The triplet wave function $\varphi_{10}$ has the form:

$$\varphi_{10} = \frac{1}{\sqrt{8}} \left\{ \xi_1(\boldsymbol{\rho},\mathbf{g}) \left[ \begin{pmatrix} 1 \\ 0 \\ 0 \\ 0 \end{pmatrix}\begin{pmatrix} 0 \\ 1 \\ 0 \\ 0 \end{pmatrix} + \begin{pmatrix} 0 \\ 1 \\ 0 \\ 0 \end{pmatrix}\begin{pmatrix} 1 \\ 0 \\ 0 \\ 0 \end{pmatrix} \right] + \xi_2(\boldsymbol{\rho},\mathbf{g}) \left[ \begin{pmatrix} 0 \\ 0 \\ 0 \\ 1 \end{pmatrix}\begin{pmatrix} 0 \\ 0 \\ 1 \\ 0 \end{pmatrix} + \begin{pmatrix} 0 \\ 0 \\ 1 \\ 0 \end{pmatrix}\begin{pmatrix} 0 \\ 0 \\ 0 \\ 1 \end{pmatrix} \right] + \right.$$

$$\left. \xi_3(\boldsymbol{\rho},\mathbf{g}) \left[ \begin{pmatrix} 0 \\ 1 \\ 0 \\ 0 \end{pmatrix}\begin{pmatrix} 0 \\ 0 \\ 1 \\ 0 \end{pmatrix} + \begin{pmatrix} 0 \\ 0 \\ 1 \\ 0 \end{pmatrix}\begin{pmatrix} 0 \\ 1 \\ 0 \\ 0 \end{pmatrix} \right] + \xi_4(\boldsymbol{\rho},\mathbf{g}) \left[ \begin{pmatrix} 1 \\ 0 \\ 0 \\ 0 \end{pmatrix}\begin{pmatrix} 0 \\ 0 \\ 0 \\ 1 \end{pmatrix} + \begin{pmatrix} 0 \\ 0 \\ 0 \\ 1 \end{pmatrix}\begin{pmatrix} 1 \\ 0 \\ 0 \\ 0 \end{pmatrix} \right] \right\}, \quad (21)$$

where the functions $\xi_i$ satisfy the same normalization ($\int |\xi_i|^2 \rho d\rho d\phi = 1$).

The triplet state $\varphi_{11}$ is defined by three normalized functions $\chi_{i=1,2,3}$:

$$\varphi_{11} = \frac{1}{2} \left\{ \chi_1(\boldsymbol{\rho},\mathbf{g}) \begin{pmatrix} 1 \\ 0 \\ 0 \\ 0 \end{pmatrix}\begin{pmatrix} 1 \\ 0 \\ 0 \\ 0 \end{pmatrix} + \chi_2(\boldsymbol{\rho},\mathbf{g}) \left[ \begin{pmatrix} 1 \\ 0 \\ 0 \\ 0 \end{pmatrix}\begin{pmatrix} 0 \\ 0 \\ 1 \\ 0 \end{pmatrix} + \begin{pmatrix} 0 \\ 0 \\ 1 \\ 0 \end{pmatrix}\begin{pmatrix} 1 \\ 0 \\ 0 \\ 0 \end{pmatrix} \right] + \chi_3(\boldsymbol{\rho},\mathbf{g}) \begin{pmatrix} 0 \\ 0 \\ 1 \\ 0 \end{pmatrix}\begin{pmatrix} 0 \\ 0 \\ 1 \\ 0 \end{pmatrix} \right\}. \quad (22)$$

The similar form is valid for the triplet state $\varphi_{1-1}$:

$$\varphi_{1-1} = \frac{1}{2}\left\{\eta_1(\boldsymbol{\rho},\mathbf{g})\begin{pmatrix}0\\1\\0\\0\end{pmatrix}\begin{pmatrix}0\\1\\0\\0\end{pmatrix} + \eta_2(\boldsymbol{\rho},\mathbf{g})\left[\begin{pmatrix}0\\1\\0\\0\end{pmatrix}\begin{pmatrix}0\\0\\0\\1\end{pmatrix} + \begin{pmatrix}0\\0\\0\\1\end{pmatrix}\begin{pmatrix}0\\1\\0\\0\end{pmatrix}\right] + \eta_3(\boldsymbol{\rho},\mathbf{g})\begin{pmatrix}0\\0\\0\\1\end{pmatrix}\begin{pmatrix}0\\0\\0\\1\end{pmatrix}\right\}, \quad (23)$$

where the three functions $\eta_{i=1,2,3}$ must also be normalized.

**7. The energies and wave functions of the massless bosons**

A closed system of equations for the functions $\xi_i (i=1,2,3,4)$ of the triplet state $\varphi_{10}$ can easily be obtained by substituting (21) in the integral equation (19). As a result, we have:

$$\xi_1 = -\alpha \int \frac{d\boldsymbol{\rho}_1}{\rho_1} \int \frac{d\mathbf{p}_\perp}{(2\pi)^2} e^{i\mathbf{p}_\perp(\boldsymbol{\rho}-\boldsymbol{\rho}_1)} \frac{(\omega^2 + 2m\omega)(\xi_1 - \xi_2) - (2mg + g\omega)(\xi_3 - \xi_4)}{\omega(\omega^2 - 4\varepsilon_p^2)}$$

$$\xi_2 = -\alpha \int \frac{d\boldsymbol{\rho}_1}{\rho_1} \int \frac{d\mathbf{p}_\perp}{(2\pi)^2} e^{i\mathbf{p}_\perp(\boldsymbol{\rho}-\boldsymbol{\rho}_1)} \frac{(-\omega^2 + 2m\omega)(\xi_1 - \xi_2) - (2mg - g\omega)(\xi_3 - \xi_4)}{\omega(\omega^2 - 4\varepsilon_p^2)}$$

$$\xi_3 = -\alpha \int \frac{d\boldsymbol{\rho}_1}{\rho_1} \int \frac{d\mathbf{p}_\perp}{(2\pi)^2} e^{i\mathbf{p}_\perp(\boldsymbol{\rho}-\boldsymbol{\rho}_1)} \frac{(-\omega^2 - g^2 + 4\varepsilon_p^2)(\xi_3 - \xi_4) + g\omega(\xi_1 - \xi_2)}{\omega(\omega^2 - 4\varepsilon_p^2)}$$

$$\xi_4 = -\xi_3. \qquad (24)$$

The first and second equations of the system (24) result in the following relation between $\xi_1$ and $\xi_2$: $(1 - \frac{\omega}{2m})\xi_1 = (1 + \frac{\omega}{2m})\xi_2$. Since these functions must be normalized, we immediately conclude that $\xi_1 = \xi_2 = 0$.

From the third and fourth equations of the system (24) we obtain:

$$\xi_3(\boldsymbol{\rho}) = \frac{2\alpha}{\omega} \int \frac{d\boldsymbol{\rho}_1}{\rho_1} \xi_3(\boldsymbol{\rho}_1) \int \frac{d\mathbf{p}_\perp}{(2\pi)^2} e^{i\mathbf{p}_\perp(\boldsymbol{\rho}-\boldsymbol{\rho}_1)} \frac{(4\varepsilon_p^2 - 2\omega^2)}{(4\varepsilon_p^2 - \omega^2)}. \qquad (25)$$

Eq. (25) is reduced to the form:

$$\left(1-\frac{2\alpha}{\omega\rho}\right)\xi_3(\boldsymbol{\rho}) = -\frac{\alpha\omega}{2}\int\frac{d\boldsymbol{\rho}_1}{\rho_1}\xi_3(\boldsymbol{\rho}_1)\int\frac{d\mathbf{p}_\perp}{(2\pi)^2}\frac{e^{i\mathbf{p}_\perp(\boldsymbol{\rho}-\boldsymbol{\rho}_1)}}{p_\perp^2+m^2}. \qquad (26)$$

The differential equation corresponding to the integral equation (26), can be easily found by action of the operator $(\Delta_\perp - m^2)$ ($\Delta_\perp = \dfrac{\partial^2}{\partial\boldsymbol{\rho}^2}$ is the two-dimensional Laplacian in variables $\rho, \phi$) on both sides of Eq. (26):

$$\left(\Delta_\perp - m^2\right)\left(1-\frac{2\alpha}{\omega\rho}\right)\xi_3(\boldsymbol{\rho}) = -\frac{\alpha\omega}{2}\int\frac{d\boldsymbol{\rho}_1}{\rho_1}\xi_3(\boldsymbol{\rho}_1)\int\frac{d\mathbf{p}_\perp}{(2\pi)^2}\frac{\left(\Delta_\perp - m^2\right)e^{i\mathbf{p}_\perp(\boldsymbol{\rho}-\boldsymbol{\rho}_1)}}{p_\perp^2+m^2}$$

Taking into account $\Delta_\perp e^{i\mathbf{p}_\perp\boldsymbol{\rho}} = -p_\perp^2 e^{i\mathbf{p}_\perp\boldsymbol{\rho}}$ and $\int\dfrac{d\mathbf{p}_\perp}{(2\pi)^2}e^{i\mathbf{p}_\perp(\boldsymbol{\rho}-\boldsymbol{\rho}_1)} = \delta(\boldsymbol{\rho}-\boldsymbol{\rho}_1)$, we find:

$$\left(\Delta_\perp - m^2\right)\tilde{\xi}_3 = \frac{1}{2}\frac{\alpha\omega}{\rho - \dfrac{2\alpha}{\omega}}\tilde{\xi}_3, \qquad (27)$$

where the function $\tilde{\xi}_3 = \left(1-\dfrac{2\alpha}{\omega\rho}\right)\xi_3$ is introduced. In the usual units Eq. (27) is written as:

$$\left(-\frac{\hbar^2}{m}\Delta_\perp + \frac{e^2\dfrac{\omega}{2mc^2}}{\rho - \dfrac{2e^2}{\omega}}\right)\tilde{\xi}_3 = -mc^2\tilde{\xi}_3.$$

Eq. (27) on the eigenfunctions $\tilde{\xi}_3$ and eigenvalues $\omega$ is formally like to the Schrodinger equation. However, the effective interaction between the fermions of the bound pair depends on the energy $\omega$. Only at negative energies this interaction is attracting. Note a strong renormalization of the electromagnetic interaction constant at large negative energies $\omega \ll -2mc^2$:

$$\alpha_{eff} = \alpha\frac{|\omega|}{2mc^2},$$

that, in fact, leads to the boson formation.

For positive energies $\omega > 0$ and $\rho > 2\alpha/\omega$ the interaction is repulsive. In addition, the wave functions $\xi_{3,4}$ can not be normalized, because, according to (27), there is a singularity of the wave function at $\rho = 2\alpha/\omega$.

In Eq. (27) it is convenient to introduce dimensionless variables: $\omega \to -\omega * m$ (it means that the boson moves backward in time) $p_\perp \to p_\perp m$, $\rho \to \rho * \lambda_c$, where $\lambda_c = \frac{1}{m}$ - the Compton wavelength of the fermion. Then Eq. (27) is rewritten as:

$$\left(\frac{\partial^2}{\partial \rho^2} + \frac{1}{\rho}\frac{\partial}{\partial \rho} + \frac{1}{\rho^2}\frac{\partial^2}{\partial \phi^2} - 1\right)\tilde{\xi}_3 = -\frac{\frac{\alpha\omega}{2}}{\rho + \frac{2\alpha}{\omega}}\tilde{\xi}_3. \qquad (28)$$

The angular dependence of the wave function is sought in the form $\tilde{\xi}_3(\phi) \propto \exp(im\phi)$, where $m = 0, \pm 1, \pm 2...$ - the quantum numbers of the z-projection of the orbital angular momentum of the transverse motion of the bound pair, $L_z = -i\frac{\partial}{\partial \phi}$. Although we use the notation of these numbers as well as the fermion mass, this should not cause confusion.

To determine the possible values of the quantum number $m$, we consider Eq. (28) for $\rho \ll \frac{2\alpha}{\omega}$. Using the substitution $\tilde{\xi}_3(\rho) = u(t)\exp(im\phi)$ with $\rho = 2t/\sqrt{\omega^2 - 4}$, reduced to the Bessel equation on $u(t)$. Given the limits of the Bessel functions $Y_{|m|}(t)$ for small values of the argument, we obtain

$$u(t) \propto \begin{cases} \ln(t), m = 0 \\ t^{-|m|}, m = \pm 1, \pm 2... \end{cases}$$

Because $\xi_3 = \rho\tilde{\xi}_3/(\rho + \frac{2\alpha}{\omega})$, we find that the normalized wave functions exist only for the three quantum numbers $m = 0, \pm 1$.

Taking into account $\tilde{\xi}_3(\rho)$ found for small $\rho$, the substitution $\tilde{\xi}_3(\rho) = \rho^{-|m|}\exp(-\rho)w(x)\exp(im\phi)$ with $x = 2\rho$ reduces Eq. (28) to the form:

$$xw''_{xx} - (x + 2|m| - 1)w'_x - \left(\frac{-2|m|+1}{2} - \frac{\alpha\omega}{4}\frac{x}{x + \frac{4\alpha}{\omega}}\right)w = 0. \qquad (29)$$

For $\rho >> \dfrac{2\alpha}{\omega}$ Eq. (29) becomes the Kummer equation and has the solution:

$$w(2\rho) = AF(\dfrac{1}{2} - |m| - \dfrac{\alpha\omega}{4}, 1 - 2|m|, 2\rho), \qquad (30)$$

where $F$ - the confluent hypergeometric function, $A$ - a normalization constant. According to (30), the normalized wave functions $\widetilde{\xi}_3(\rho)$ can be only at

$$-|m| + \dfrac{1}{2} - \dfrac{\alpha\omega}{4} = -n,$$

where $n = 0,1,2,3...$. In this case, the functions (30) are the Laguerre polynomials $L_n^{(-2|m|)}(2\rho)$.

Thus, in the triplet state $\varphi_{10}$ the high-energy branch of the massless bosons has been found with the normalized wave functions:

$$\varphi_{10}^{nm} = \xi_3(\rho,\omega_{nm})e^{im\phi}e^{i\omega_{nm}t + ig_{nm}z}\left\{\begin{pmatrix}0\\1\\0\\0\end{pmatrix}\begin{pmatrix}0\\0\\1\\0\end{pmatrix} + \begin{pmatrix}0\\0\\1\\0\end{pmatrix}\begin{pmatrix}0\\1\\0\\0\end{pmatrix} - \begin{pmatrix}1\\0\\0\\0\end{pmatrix}\begin{pmatrix}0\\0\\0\\1\end{pmatrix} - \begin{pmatrix}0\\0\\0\\1\end{pmatrix}\begin{pmatrix}1\\0\\0\\0\end{pmatrix}\right\},$$

where

$$\xi_3 = \begin{cases} A\rho^{-|m|}e^{-\rho}L_n^{(-2|m|)}(2\rho), \rho >> \dfrac{2\alpha}{\omega_{nm}} \\ B\rho u\left(\dfrac{\rho}{2}\sqrt{\omega_{nm}^2 - 4}\right), \rho << \dfrac{2\alpha}{\omega_{nm}} \end{cases}.$$

The quantized energies of the composite boson (in usual units) are $\omega_{nm} = \dfrac{4}{\alpha}mc^2\left(n - |m| + \dfrac{1}{2}\right)$. Here $m = 0$ in the case $n = 0$ and $m = 0, \pm 1$ at $n \geq 1$. The boson momentum $g_{nm} = \omega_{nm}/c$ is also quantized. For the states $\varphi_{10}^{n0}$ the z-projection of the orbital angular momentum of the transverse motion of the bound pair $L_z(m = 0) = 0$, for the states $\varphi_{10}^{n1}$ this projection $L_z(m = 1) = \hbar$ and for $\varphi_{10}^{n,-1}$ - $L_z(m = -1) = -\hbar$. The value $n - |m| = 0$ corresponds to the lower limit of the boson spectrum, $\omega_h = \dfrac{2}{\alpha}mc^2$. That is, the bosons are highly energetic.

The average two-dimension radius of the wave functions of the transverse motion is:

$$\overline{\rho}_{nm} = \int \rho \, |\varphi_{10}^{nm}|^2 \, d\rho$$

It turned out that the average radius of the wave functions increases with the increasing energy $\omega_{nm}$ (or the quantum number $n$). Thus, $\bar{\rho}_{00} = 2\lambda_C, \bar{\rho}_{10} = 3.23\lambda_C$ and $\bar{\rho}_{20} = 4.8\lambda_C$ at $m=0$, $\bar{\rho}_{11} = 1.73\lambda_C, \bar{\rho}_{21} = 3\lambda_C$ and $\bar{\rho}_{31} = 4.6\lambda_C$ at $|m|=1$.

The triplet states $\varphi_{11}$ and $\varphi_{1-1}$ are interrelated since they are determined by a common system of equations. The system is:

$$\chi_1 = -\alpha \int \frac{d\boldsymbol{\rho}_1}{\rho_1} \int \frac{d\mathbf{p}_\perp}{(2\pi)^2} e^{i\mathbf{p}_\perp(\boldsymbol{\rho}-\boldsymbol{\rho}_1)} \frac{\left(\omega^2 - 2\varepsilon_p^2 + 2m\omega - \frac{g^2}{2} + 2m^2\right)(\chi_1 - \chi_3) + 2(\hat{p}_x - i\hat{p}_y)^2(\eta_1 - \eta_3)}{\omega\left(\omega^2 - 4\varepsilon_p^2\right)}$$

$$\chi_2(\boldsymbol{\rho}) = -2mg\alpha \int \frac{d\boldsymbol{\rho}_1}{\rho_1} \int \frac{d\mathbf{p}_\perp}{(2\pi)^2} e^{i\mathbf{p}_\perp(\boldsymbol{\rho}-\boldsymbol{\rho}_1)} \frac{(\chi_1(\boldsymbol{\rho}_1) - \chi_3(\boldsymbol{\rho}_1))}{\omega\left(\omega^2 - 4\varepsilon_p^2\right)}$$

$$\chi_3 = +\alpha \int \frac{d\boldsymbol{\rho}_1}{\rho_1} \int \frac{d\mathbf{p}_\perp}{(2\pi)^2} e^{i\mathbf{p}_\perp(\boldsymbol{\rho}-\boldsymbol{\rho}_1)} \frac{\left(\omega^2 - 2\varepsilon_p^2 - 2m\omega - \frac{g^2}{2} + 2m^2\right)(\chi_1 - \chi_3) + 2(\hat{p}_x - i\hat{p}_y)^2(\eta_1 - \eta_3)}{\omega\left(\omega^2 - 4\varepsilon_p^2\right)}$$

$$\eta_1 = \chi_1^*$$
$$\eta_2 = -\chi_2^*$$
$$\eta_3 = \chi_3^*. \qquad (31)$$

From the system (31) we can see that the mixed triplet state $\varphi_{11} + \varphi_{1-1}$ is possible if the following system of equations:

$$\chi_1 + \chi_3 = \alpha \int \frac{d\boldsymbol{\rho}_1}{\rho_1} \int \frac{d\mathbf{p}_\perp}{(2\pi)^2} e^{i\mathbf{p}_\perp(\boldsymbol{\rho}-\boldsymbol{\rho}_1)} \frac{\chi_1 - \chi_3}{p_\perp^2 + m^2}$$

$$\chi_1 - \chi_3 = -\frac{\alpha}{\omega} \int \frac{d\boldsymbol{\rho}_1}{\rho_1} \int \frac{d\mathbf{p}_\perp}{(2\pi)^2} e^{i\mathbf{p}_\perp(\boldsymbol{\rho}-\boldsymbol{\rho}_1)} \frac{p_\perp^2(\chi_1 - \chi_3) - (\hat{p}_x - i\hat{p}_y)^2(\chi_1^* - \chi_3^*)}{p_\perp^2 + m^2} \qquad (32)$$

has the normalized solutions for the functions $\chi_{1,2,3}$. We could not solve this system for the mixed triplet state.

For the singlet state $\varphi_{00}$ (20) the closed system of integral equations is:

$$\theta_1 + \theta_2 = \frac{\alpha}{\omega} \int \frac{d\rho_1}{\rho_1} \int \frac{d\mathbf{p}_\perp}{(2\pi)^2} e^{i\mathbf{p}_\perp(\rho-\rho_1)} \frac{(\omega^2 - g^2)(\theta_1 + \theta_2) - m\omega(\theta_1 - \theta_2) + mg(\theta_3 - \theta_4)}{p_\perp^2 + m^2}$$

$$\theta_1 - \theta_2 = \frac{\alpha}{2\omega} \int \frac{d\rho_1}{\rho_1} \int \frac{d\mathbf{p}_\perp}{(2\pi)^2} e^{i\mathbf{p}_\perp(\rho-\rho_1)} \frac{(4p_\perp^2 - \omega^2)(\theta_1 - \theta_2) + 4m\omega(\theta_1 + \theta_2) + g\omega(\theta_3 - \theta_4)}{p_\perp^2 + m^2}$$

$$\theta_3 + \theta_4 = -\frac{\alpha}{\omega} \int \frac{d\rho_1}{\rho_1} \int \frac{d\mathbf{p}_\perp}{(2\pi)^2} e^{i\mathbf{p}_\perp(\rho-\rho_1)} \frac{(4m^2 - \omega^2 + g^2)(\theta_3 + \theta_4)}{p_\perp^2 + m^2}$$

$$\theta_3 - \theta_4 = -\frac{\alpha}{2\omega} \int \frac{d\rho_1}{\rho_1} \int \frac{d\mathbf{p}_\perp}{(2\pi)^2} e^{i\mathbf{p}_\perp(\rho-\rho_1)} \frac{(4m^2 - \omega^2)(\theta_3 - \theta_4) - 4mg(\theta_1 + \theta_2) + g\omega(\theta_1 - \theta_2)}{p_\perp^2 + m^2}. \quad (33)$$

Introducing dimensionless variables $\omega \to \omega * m$, $p_\perp \to p_\perp m$, $\rho \to \rho * \lambda_c$ and $g = \beta\omega$, where $\beta = +1$ for the moving of the singlet boson forward in time and $\beta = -1$ - the moving backward, the system (33) is rewritten as:

$$\theta_1 + \theta_2 = -\alpha \int \frac{d\rho_1}{\rho_1} \int \frac{d\mathbf{p}_\perp}{(2\pi)^2} e^{i\mathbf{p}_\perp(\rho-\rho_1)} \frac{\theta_1 - \theta_2 - \beta(\theta_3 - \theta_4)}{p_\perp^2 + 1}$$

$$\theta_1 - \theta_2 = \frac{2\alpha}{\omega} \int \frac{d\rho_1}{\rho_1} \int \frac{d\mathbf{p}_\perp}{(2\pi)^2} e^{i\mathbf{p}_\perp(\rho-\rho_1)} \frac{\left(p_\perp^2 - \frac{\omega^2}{4}\right)(\theta_1 - \theta_2) + \omega(\theta_1 + \theta_2) + \beta\frac{\omega^2}{4}(\theta_3 - \theta_4)}{p_\perp^2 + 1}$$

$$\theta_3 + \theta_4 = -\frac{4\alpha}{\omega} \int \frac{d\rho_1}{\rho_1} \int \frac{d\mathbf{p}_\perp}{(2\pi)^2} e^{i\mathbf{p}_\perp(\rho-\rho_1)} \frac{\theta_3 + \theta_4}{p_\perp^2 + 1}$$

$$\theta_3 - \theta_4 = -\frac{2\alpha}{\omega} \int \frac{d\rho_1}{\rho_1} \int \frac{d\mathbf{p}_\perp}{(2\pi)^2} e^{i\mathbf{p}_\perp(\rho-\rho_1)} \frac{\left(1 - \frac{\omega^2}{4}\right)(\theta_3 - \theta_4) - \beta\omega(\theta_1 + \theta_2) + \beta\frac{\omega^2}{4}(\theta_1 - \theta_2)}{p_\perp^2 + 1}. \quad (34)$$

The second equation of (34) is reduced to the form:

$$\left(1 - \frac{2\alpha}{\omega\rho}\right)(\theta_1 - \theta_2) = -\frac{2\alpha}{\omega} \int \frac{d\rho_1}{\rho_1} \int \frac{d\mathbf{p}_\perp}{(2\pi)^2} e^{i\mathbf{p}_\perp(\rho-\rho_1)} \frac{\left(1 + \frac{\omega^2}{4}\right)(\theta_1 - \theta_2) - \omega(\theta_1 + \theta_2) - \beta\frac{\omega^2}{4}(\theta_3 - \theta_4)}{p_\perp^2 + 1}$$

From the latter equation and the first and fourth equations of the system (34) we obtain:

$$\beta(\theta_3 - \theta_4) = \left(1 - \frac{2}{\omega} - \frac{2\alpha}{\omega\rho}\right)\theta_1 - \left(1 + \frac{2}{\omega} - \frac{2\alpha}{\omega\rho}\right)\theta_2. \quad (35)$$

Substituting (35) in the first and second equations of the system (34), we find a closed system of integral equations for the two functions $\theta_{1,2}$:

$$\theta_1(\boldsymbol{\rho})+\theta_2(\boldsymbol{\rho}) = -\frac{2\alpha}{\omega}\int\frac{d\boldsymbol{\rho}_1}{\rho_1}\left[\theta_1(\boldsymbol{\rho}_1)+\theta_2(\boldsymbol{\rho}_1)+\frac{\alpha}{\rho_1}(\theta_1(\boldsymbol{\rho}_1)-\theta_2(\boldsymbol{\rho}_1))\right]\int\frac{d\mathbf{p}_\perp}{(2\pi)^2}\frac{e^{i\mathbf{p}_\perp(\boldsymbol{\rho}-\boldsymbol{\rho}_1)}}{p_\perp^2+1}$$

$$\left(1-\frac{2\alpha}{\omega\rho}\right)(\theta_1-\theta_2) = -\frac{2\alpha}{\omega}\int\frac{d\boldsymbol{\rho}_1}{\rho_1}\left[\left(1+\frac{\alpha\omega}{2\rho}\right)(\theta_1-\theta_2)-\frac{\omega}{2}(\theta_1+\theta_2)\right]\int\frac{d\mathbf{p}_\perp}{(2\pi)^2}\frac{e^{i\mathbf{p}_\perp(\boldsymbol{\rho}-\boldsymbol{\rho}_1)}}{p_\perp^2+1}. \quad (36)$$

First we consider the case of positive energies ($\omega > 0$) of the composite boson. According to the second equation of the system (36) the function $\theta_1 - \theta_2$ has a singularity at the value $\rho = 2\alpha/\omega$. Therefore, in order to the functions were normalized, it is necessary that $\theta_1 = \theta_2$. Then the analysis of the system (36) leads to the result $\theta_1 = \theta_2 = 0$. From (35) we obtain $\theta_3 = \theta_4$, and the third equation (34) leads to the following integral equation for these functions:

$$\theta_3(\boldsymbol{\rho}) = -\frac{4\alpha}{\omega}\int\frac{d\boldsymbol{\rho}_1}{\rho_1}\theta_3(\boldsymbol{\rho}_1)\int\frac{d\mathbf{p}_\perp}{(2\pi)^2}\frac{e^{i\mathbf{p}_\perp(\boldsymbol{\rho}-\boldsymbol{\rho}_1)}}{p_\perp^2+1}. \quad (37)$$

The integral equation (37) corresponds to the following differential equation:

$$(\Delta_\perp - 1)\theta_3 = \frac{4\alpha}{\rho\omega}\theta_3. \quad (38)$$

Thus, for positive energies the functions $\theta_3 = \theta_4$ defined by Eq. (38), can only be nonzero. The study of this equation is held over.

Now we consider the case of negative energies. If the functions $\theta_3$ и $\theta_4$ are equal to each other ($\theta_3 = \theta_4$), then, in accordance with (35) the functions $\theta_1$ and $\theta_2$ can not be normalized and, therefore, $\theta_1 = \theta_2 = 0$. From the first and fourth equations of the system (34) we obtain $\theta_3 = \theta_4$ and, using the third integral equation of (34), the differential equation (38) for $\theta_3$.

If $\theta_3 \neq \theta_4$ then the closed system of integral equations (36) defining the functions $\theta_1$ и $\theta_2$, is reduced to the following system of differential equations:

$$(\Delta_\perp - 1)\theta_+ = \frac{2\alpha}{\rho\omega}\left[\theta_+ + \frac{\alpha}{\rho - \frac{2\alpha}{\omega}}\theta_-\right]$$

$$(\Delta_\perp - 1)\theta_- = \frac{2\alpha}{\rho\omega}\left[-\frac{\omega}{2}\theta_+ + \frac{\rho+\frac{\alpha\omega}{2}}{\rho - \frac{2\alpha}{\omega}}\theta_-\right]. \quad (39)$$

Here the notations $\theta_+ = \theta_1 + \theta_2$ and $\theta_- = \left(1 - \dfrac{2\alpha}{\rho\omega}\right)(\theta_1 - \theta_2)$ are introduced.

The system (39) was solved numerically. Its solution defines the functions $\theta_1$ and $\theta_2$. Then, from Eq. (38) with replacement $\theta_3 \to \theta_3 + \theta_4$ and Eq. (35) the functions $\theta_3$ and $\theta_4$ are determined. It turned out that it is possible to obtain the normalized solutions for the functions $\theta_1$ and $\theta_2$. However, the functions $\theta_3$ and $\theta_4$ are not normalized.

Thus, the singlet state $\psi_{00}$ of the massless boson is determined the wave function $\theta_3$ ($\theta_4 = \theta_3$) satisfying the Eq. (38). In the usual units it is:

$$\left(-\dfrac{\hbar^2}{m}\Delta_\perp + \dfrac{e^2}{\rho} * \dfrac{4mc^2}{\omega}\right)\theta_3 = -mc^2\theta_3.$$

This equation is formally like to the Schrodinger equation. However it eigenvalues $\omega$ determine the effective interaction potential. There is a strong renormalization of the constants of the electromagnetic interaction

$$\alpha_{eff} = \alpha\dfrac{4mc^2}{|\omega|}$$

at low energies $\omega$. For positive energies the interaction potential is repulsive and, therefore, the boson solutions do not exist (that is, $\theta_3 = 0$). At negative energies ($\omega < 0$) the potential energy is attractive, and the strongly coupled composite particle can exist.

Eq. (38) with the replacement $\omega \to -\omega$ (the move backward in time) is rewritten as:

$$\left(\dfrac{\partial^2}{\partial\rho^2} + \dfrac{1}{\rho}\dfrac{\partial}{\partial\rho} + \dfrac{1}{\rho^2}\dfrac{\partial^2}{\partial\phi^2} - 1\right)\theta_3 = -\dfrac{4\alpha}{\omega\rho}\theta_3. \qquad (40)$$

Using the substitution $\theta_3(\rho) = \rho^{|m|}\exp(-\rho)w(x)\exp(im\phi)$, where $x = 2\rho$  $m = 0, \pm1,..$ - the quantum numbers of the z-projection of the orbital angular momentum of the transverse motion of the bound pair, Eq. (40) is reduced to the Kummer function :

$$xw''_{xx} - (x - 2|m| - 1)w'_x - \left(\dfrac{2|m|+1}{2} - \dfrac{2\alpha}{\omega}\right)w = 0.$$

Its solution is:

$$w(2\rho) = AF(|m| + \dfrac{1}{2} - \dfrac{2\alpha}{\omega}, 2|m|+1, 2\rho)$$

The wave functions $\theta_3$ are normalized only at

$$|m| + \frac{1}{2} - \frac{2\alpha}{\omega} = -n,$$

where $n = 0,1,2,3...$. In these cases the functions $w(2\rho)$ are the Laguerre polynomials $L_n^{(2|m|)}(2\rho)$.

Thus, in the singlet state the low-energy branch of the massless bosons has been found with the normalized wave functions:

$$\varphi_{00}^{nm}(\boldsymbol{\rho},z) = A\rho^{|m|}e^{-\rho}L_n^{(2|m|)}(2\rho)e^{im\phi}e^{i\omega_{nm}t+ig_{nm}z}\left[\begin{pmatrix}0\\0\\1\\0\end{pmatrix}\begin{pmatrix}0\\1\\0\\0\end{pmatrix} - \begin{pmatrix}0\\1\\0\\0\end{pmatrix}\begin{pmatrix}0\\0\\1\\0\end{pmatrix} + \begin{pmatrix}1\\0\\0\\0\end{pmatrix}\begin{pmatrix}0\\0\\0\\1\end{pmatrix} - \begin{pmatrix}0\\0\\0\\1\end{pmatrix}\begin{pmatrix}1\\0\\0\\0\end{pmatrix}\right].$$

and the energies $\omega_{nm} = \dfrac{2\alpha mc^2}{n+|m|+\dfrac{1}{2}}$, where $n = 0,1,2,3...$ and $m = 0,\pm1,\pm2...$.

The top limit of this low-energy branch equals to $\omega_{0,0} = 4\alpha mc^2$. The momentum of the boson $g_{nm} = \omega_{nm}/c$ is also quantized.

The average two-dimensional radius of the transverse wave functions is:

$$\overline{\rho}_{00}^{nm} = 2\pi \int_0^\infty \rho^2 |\varphi_{00}^{nm}|^2 d\rho.$$

The change of the average radius of the wave functions $\varphi_{00}^{0m}$ (the quantum number $n = 0$) is due to the orbital angular momentum of the transverse motion of the bound pair. In this case we obtain $\overline{\rho}_{00}^{0m} = (|m|+1)\lambda_C$. For the wave functions $\varphi_{00}^{n0}$ the average transverse radius increases with the increasing of the momentum of the boson. Thus, $\overline{\rho}_{00}^{10} = 2.33\lambda_C$ and $\overline{\rho}_{00}^{20} = 3.8\lambda_C$.

**8. Conclusions**

The results presented above were obtained at the neglecting the interaction retardation between fermions of the bound pair. For the states of the forms (20)-(23) we can find the integral equation for the eigenfunctions and energy eigenvalues of the composite boson with account for the retardation. However, we could not solve these equations.

It should be discussed what fermions can these strongly coupled wave states exist. In the standard QED model the theory presented above can not be applied to the two-particle system of electron and positron, since the model is based on the hole theory of positrons [15]. In this theory, instead of (2), the other solution of Eq. (1) is used for the free electron-positron propagator, and the states with negative energies are not available for electrons in any scattering processes [15]. However, these positive energy states do not form the complete basis of the eigenstates of the Dirac equation [17]. Not any wave function can be presented as the superposition of the states of the incomplete basis.

Since no evidence of the hole concept for positron are available in the modern theory of antiparticles [18], only the experiment can either confirm or disprove the existence of these massless bosons for the electron-positron system. We would like to discuss the following experiment in which these strongly bound states, if they exist, can be detected.

These composite bosons are a hidden matter because the masses of electron and positron are disappeared in the strongly bound states. For the low-energy singlet bosons one can try to distinguish the annihilation channel of parapositronium

$$(ee^+)_{1^1S_0} \to 2\gamma ,$$

where the sum of the photon momentums $\mathbf{k}_1 + \mathbf{k}_2 = 0$, from the next predicted reaction

$$(ee^+)_{1^1S_0} \to g + 2\gamma ,$$

where $\mathbf{g} + \mathbf{k}_1 + \mathbf{k}_2 = 0$, $\mathbf{g}$ - the momentum of the massless boson ($g$). The low-energy boson branch has the upper limit equal to $4\alpha mc^2 = 14.6$ keV that is very small as compared with the electron mass. Nevertheless the deviation angle between almost the opposite directions of the two momentums of the simultaneously detected photons will be approximately $1^0$. Therefore, a high experimental resolution is needed for the detection of the latter reaction. Finding the photon energies and the deviation angle, it is possible to obtain experimentally the momentum dependence of the boson energy and to compare with the theoretical dependence $\omega = cg$.

For high energy bosons the lower limit of the energy spectrum is 140 MeV. Their electromagnetic interaction with any charged particle $A$

$$A + g \to A' + e^- + e^+$$

can lead to the destruction of the triplet bosons and the formation of excess free electrons and positrons.

**References**

1. J.S. Goldstein, Phys. Rev. **91**, 1516 (1953).

2. E.E. Salpeter and H.A. Bethe, Phys. Rev. **84**, 1232 (1951).

3. R. Delbourgo, A. Salam, J. Strathdee, Nuovo Cim. **50**, 193 (1967)

4. L.G. Suttorp, Nuovo Cim. A **35**, 257 (1976)

5. M. Bawin, J. Phys. A: Math. Gen. **10**, 2037 (1977)

6. G. Feldman, T. Fulton, Phys. Rev. D **8**, 3616 (1973)

7. G.B. Mainland, J. Math. Phys. **27**, 1344 (1986)

8. J. Pankaj, A.J. Sommerer, D.W. Mckay *et al*., Phys. Rev. D **46**, 4029 (1992)

9. G.B. Mainland, J. Math. Phys. **27**, 1344 (1986)

10. N.E. Ligterink, A. Weber, Few-Body Syst. **46**, 115 (2009)

11. S. Ahlig, R. Alkofer, Annals Phys. **275**, 113 (1999)

12. J. Bijtebier, Nucl. Phys. A**623**, 498 (1997)

13. N. Nakanishi, in: Proc. Int. Symp. on Extended Objects and Bound Systems, Karuizawa, Japan, World Scientific, Singapore, 1992, p. 109.

14. N. Seto, Prog. Theor. Phys. (Suppl.) **95**, 25 (1988).

15. R.P. Feynman, Phys. Rev. **76**, 749 (1949).

16. A.A. Logunov and A.N. Tavkhelidze, Nuovo Cim., **29**, 380 (1963).

17. A. I. Akhiezer and V. B. Berestetskii. *Quantum Electrodynamics*. New York: Interscience Publishers, 1965.

18. A. Sudberry, *Quantum mechanics and the particles of nature*, Cambridge Univ. Press, 1986.